# A New Assumed Interaction.
# Experiments and Manifestations in Astrophysics.

Baurov Yu.A.

*Central Research Institute of Machine Building,*

*141070, Pionerskaya 4, Korolyov, Moscow region, Russia.*

**Results of experimental investigations of a new assumed interaction in nature with the aid of high-current magnets, torsion and piezoresonance balances, high-precision gravimeter, fluctuations in intensity of β-decay of radioactive elements, plasma devices and manifestations in astrophysics are presented. A possible explanation of the results obtained based on a hypothesis of global anisotropy of physical space caused by the existence of a cosmological vectorial potential $A_g$, is given. It is shown that the vector $A_g$ has the following coordinates in the second equatorial coordinate system: right ascension α = 293° ± 10°, declination δ = 36°± 10°.**

Up to now it was assumed [1,2] that there exist only four interactions (four forces) between objects in nature: the electromagnetic interaction determining by almost 99 percent the structure of all substances we have to deal with; the gravitational interaction, the most weak of all, it determines motion of massive bodies (planets around the Sun or, for example, a motor-car, since, if the latter had no weight, there would be no frictional force with the aid of which it moves); the strong interaction determining stability of all atomic nuclei (without this interaction they would fall to pieces by the action of Coulomb forces between protons); and the weak interaction that destroys many instable elementary particles, for example, neutrons. From the forces listed, the former three have a property of isotropicity, that is, their magnitude does not depend on the chosen direction in the physical space. Only the weak interaction manifests its anisotropy in local volumes on the order of $10^{-17}$cm. That is, if we cool, for example, atoms of $Co^{60}$ down to very low temperatures (1°K) and draw up spins of its nuclei along some spatial direction, the number of electrons flying out of nuclei of that element along said direction, will be considerably more than in the opposite and transversal directions. Such an experiment carried out firstly by Madame Wo, is a classical experiment in the nuclear physics [1].

The new interaction predicted by the author [3-22], was bound to have just such anisotropic properties. As will be shown below, the anisotropic properties of new interaction manifest themselves, as distinct from weak interaction, of the scales of the latter ($10^{-17}$cm) up to dimensions of our Galaxy ($10^{27}$cm) and possibly even farther. Let us explain the essence of new assumed interaction. According to the theory of byuons [17-20], masses of all elementary particles are proportional to the modulus of some summary potential $A_\Sigma$ that contains potentials of all known fields. The value of $A_\Sigma$ cannot be larger than the modulus of $A_g$ - a new fundamental, vectorial constant appearing in the definition of byuons not interacting.

Therefore an idea has arisen to act upon the process of formation of elementary particles' masses with the aid of a vectorial potential of the electromagnetic field of current systems which potential, in proximity of the current, is always directed in the direction of this current. The new force was assumed to reject substance out of a region of weakened potential to the side of $A_g$.

In the present paper, we describe only the results of experimental investigations into the properties of the assumed new force performed on the best experimental sets in the USSR and Russia, as well as some astrophysical manifestations of this force.

**Experiments:**

**Initial experiments on investigation of new force using high-current magnets, torsion and piezoresonance balances**

In this paper we consider only one experiment performed in the Institute of General Physics of RAS (IGPRAS) (director Nobel Prize winner A.M. Prokhorov) March 02, 1990 [4,5,17,18] and January 9-10, 1992 [6,7,17,18]. For investigations, the strongest resistive electromagnet in Moscow, with the field up to 15T, was used. To study the new force, a special torsion balance was manufactured.

It was arranged in a non-magnetizable metallic tube which was evacuated in order for convective flows of air do not influence the experimental results. As a test weight, a piece of beta-tin, very weak paramagnetic (that is, substance very poorly attracted by magnetic field), was used. A metallic tube with the balance was arranged in the opening of a cylindrical magnet so that the action of the magnet upon the weight was minimum. The tube could be rotated together with the weight in the opening of magnet for the weight to be placed in various opening sectors of the magnet.

The procedure of measurements was the following. By rotating the tube, we moved the test weight into regions 1,2,3,4 of the magnet (Fig.1).

In each region, an angular deviation *x* of the weight when increasing the magnitude of magnetic field up to 14T, was measured. In Fig.1 the results of the experiments are shown. As is seen from the figure, the weight was attracted to the wall of the solenoid practically identically in regions 1,2,4, but the attraction was considerably weaker in the region 3 as if some force repels the weight from the wall. After six hours, the weight in region 3 was attracted as before in regions 1,2,4, and the zone of repulsion was moved into an other region. The general conclusion from the run of experiments with high-current magnets was that the zone of repulsion moved inside the solenoid with a velocity of (15-18) degrees per hour.

The main drawbacks of this experiment and of the experiments described in [4,5,17,18] were the following ones. First, to measure the new force, the vertical of the torsion balance was to be thoroughly adjusted what could be made only with certain accuracy. Second, by the calibration of the balance was not very precise, too. Third, the used torsion balance was not able to hold a constant stretching of the filament within many hours, which was necessary in daily experiments for determining cyclicity of the new force. A balance free from the defects indicated is the quartz piezoresonance balance functioning on highly noise immune piezoresonance method of measurement, based on changing the frequency of a piezoquartz resonator under the action of the load. The measurements of the force using piezoresonance method [6,7,17,18] have convincingly repeated the results obtained with the aid of torsion balance.

The magnitude of the new discovered force was 0.01-0.07g in this experiments. The weight mass was about of 30g.

In the experiments of 1987-1994 [4-7,17,18], one of the coordinates of vector $\mathbf{A}_g$ direction in the second equatorial coordinate system was firstly determined: the right ascension $\alpha \approx 270°$. An analysis of great run of experiments with strong magnets and torsion or piezoresonance balances carried out throughout many years, has shown that the new force has of a nonlinear and nonlocal character and can be represented as a complex series (sum) in terms of changes in this summary potential $A_\Sigma$. The first term of the series is

$$F \sim N * \Delta A_\Sigma \frac{\partial \Delta A_\Sigma}{\partial x} \tag{1}$$

where N is the number of stable particles (electrons, protons, and neutrons) in the test body; $\Delta A_\Sigma$ equal to the difference in changes of the summary potential $|\vec{A}_\Sigma|$ at the location points of a test body and sensor element; $x$ - length of an arc of a circle for this experiments (a space coordinate).

**Experimental investigations of new force with the aid of high′- precision gravimeter.**

Let us begin with the gravimeters, that is devices intended for measurements of gravity oscillations. On Earth the variations of gravitational field are determined mainly by the Moon and the Sun, - i.e. by so-called lunar and solar tides.

Describe the first run of experiments in 1994-1996 [17,18].

As an experimental installation, a tide gravimeter created at the Scternberg State Astronomic Institute of the Moscow State University (SSAI MSU) on the basis of Canadian standard quartz gravimeter "Sodin", was used. The experiments were performed at a specialized gravimetric laboratory located in a special box 2m below grade. The foundation of gravimeter was separated from that of the building.

The principal diagram of the quartz sensing system is shown in [16-18]. Its basis was a quartz lever 2cm in length with a platinum mass $m$ = 0.05g. The accuracy of minute's data was $(0.5-1.2) \cdot 10^{-8}$ms$^{-2}$. In the vicinity of the gravimeter a magnet was arranged as an amplifier of new force.

Consider one result. The fixed event of 18.03.1996 at 20.54 had huge amplitude of 15.2 moon tides with an interval $\Delta t \approx 10$min. In Fig.2 shown is a time profile of this event. It is seen from the figure that during ~ 10 minutes, the Earth's gravitational pull on the platinum weight of the gravimeter was diminished and then the reading returned to the curve corresponding to the moon tide.

In the run of experiments in 1996 year, 12 events were recorded in all. In 11 cases, the gravimeter's axis of sensibility (being normal to the surface of the Earth) pointed to the side of vector $A_g$, the direction of which had been estimated earlier in experiments with strong magnets.

To eliminate possible systematic errors, a new run of experiments was carried out in 1999-2001 [16,18]. There was used a method of two gravimeters one of which had an attached magnet. The gravimeters were arranged at underground (at a depth of ~ 10m) gravimetric laboratory of SSAI MSU restored on a special one-piece foundation-pedestal separated from the foundation of the main building.

The measurements of the force using the method of two gravimeters [16,18] have reliably confirmed the results obtained with the aid of single gravimeter.

Thus the experiments carried out with the gravimeter have shown that after some refinement (attachment of a magnet), a standard device gains the capacity to catch signals of new nature.

**Experimental investigations of new force through measuring fluctuations in intensity of β-decay of radioactive elements.**

It is known from the traditional physics that the β-decay is 100 percent random process being influenced neither by enormous magnetic fields, no by huge pressures, no by any other known factors [24,25]. But the author has assumed even in his earlier work [3] that the β-decay rate of radioactive elements can be affected by the vector potential of electromagnetic field. Experimental investigations of changes in the rate of the decay of radioactive elements using the new force date back to the mid-1990s [8,17,18]. In this paper we show only the part of our last works [9,10,18]. In this works the results of our experiment performed at the same time (simultaneously), during some month of 2000, at Institute for Nuclear Research, Russian Academy of Sciences of (INR RAS, Troitsk, Moscow region) and Joint Institute for Nuclear Research (JINR, Dubna, Moscow region) with the use of germanium-lithium detectors, are given. The goal of this experiment was to measure changes in the β-decay rate of radioactive elements, for higher plausibility of results, in parallel at different points of Earth and over the most prolonged period of time as possible.

The towns Dubna and Troitsk are nearly at the same meridian (that of Moscow) ~ 140 km apart. The experiment went on from the middle of February, 2000, till the middle of May, 2000. Because of complexity and large amount of material on the subject we give here only final results the experiment at Troisk.

In Fig.3 shown are the obtained data on fluxes of γ-quanta accompanying the β-decay in the experiments with $^{60}$Co over the period from 16.24 of Moscow time at March 15, 2000, till April 10, 2000, inclusive. Here changes in γ-quanta count rate reached 1 percent relative to statistical average over about two-day's time period. At Dubna ($^{137}$Cs), the similar values were no more than 0.2% of statistical average over the same time period. In the works [9,10,18] is shown that the obtained result cannot be explained by known physical phenomena. Therefore it is said about a new "cosmological" factor influencing the process of β-decay of nuclei and being connected with the new assumed interaction of objects in nature caused by the existence of cosmological vector potential **A**$_g$.

In Figs. 4 the small circles indicate places of observation of maximum fluxes of γ-quanta during the $^{60}$Co β-decay at Troitsk. The numbered maximums Figs 3 are corresponded by numbers of arrows in Figs.4 drawn from place, where the extremum was observed, along the tangent to the Earth's parallel of latitude along which the vector potential of the terrestrial magnetic dipole is directed.

As is seen from Fig.4 the entire set of numbered arrows can be clearly subdivided, as to their directions, into three subsets with an accuracy of ±10%. In Fig.4 we see subsets $T_1$ (1, 2, 3, 4, 5, 6, 7, 10, 11, 15, 17, 18, 21); $T_2$ (8, 9, 19, 20); $T_3$ (13, 22, 12, 16). In this experiment we have seen the same pattern of behavior of maximums and minimums of β-decay rate for different elements $^{137}$Cs and $^{60}$Co, in the same interval of time at different experimental spaced 140 km apart. Therewith the direction of arrows of subset $T_2$ gives the direction of vector **A**$_g$ equal to ≈ 285°±15°. Thus the experiment considered has qualitatively confirmed the results of earlier experiments (see [4-7, 17,18]) on investigating changes in the β-decay rate of radioactive elements and refined the direction of **A**$_g$.

It should be noted that periodic (~ 24 hour's) variations in fluxes of particles during α- and β-decays of radioactive elements were observed in the work [26], too. The analysis of its results also points to the existence of above mentioned property of physical space which manifests itself in α- and β-decays. In the work [27] when scanning the celestial sphere by a reflecting telescope with a radioactive source $^{60}$Co at the focus, bursts in the count rate were registered. The directions of tangent lines to the to the Earth's parallels at the time points of bursts in $^{60}$Co activity are the same subsets $T_1$, $T_2$, $T_3$.

At the begin of paragraph we mentioned that the β-decay was considered to date as a 100% randon process. But there are experiments other than the above mentioned which also indicate that the β-decay of $^7$Be, for example, can be influenced by introducing it to the crystal lattice of the fulleren $^{60}$C which leads to the change[28] in the intensity of β-decay up to 0.83% as compared with the metallic Be.

**Experimental Investigations of the new assumed force using plasma devices.**

In Refs. [18,21], the detailed description of the investigation of the new force with the aid three different plasma devices in the different institute is shown. In this paper we are show the experimental investigations were performed in a laboratory of the Physics Department of Moscow State University named by M.V.Lomonosov [18,21,22].

The main parameter measured in the experiment was the deflection of the beam of the mirror-galvanometer oscillograph. This deflection, proportional to the radiative intensity of plasma, was recorded on the photographic strip and gave information on the amount of energy released in the discharge of the plasma generator and, while scanning the celestial sphere, on the direction of maximum action of the new force upon the particles of the plasma discharge.

To investigate the direction of the global anisotropy of physical space caused by the vector $A_g$, as well as the new interaction connected with this vector, the following technique was used. In the first of experiments (15.12.1999-3.05.2000), the plasma generator was rotated only around the vertical axis. The start time of the experiment was determined by the position of the vector $A_g$ near the horizontal plane.

From 15.12.1999 till 3.05.2000 32 experiments with rotation of the plasma generator around the vertical axis through 360° were carried out. The duration of one experiment (25 or 30 shots) was no more than 30-25 min. As an illustration, in Fig.5 the values of deflection L of the beam of the mirror-galvanometer oscillograph (in millimeters) in dependence on the angle of rotation θ in the experiments 15.12.1999 and 20.01.2000 are shown.

In the experiment of 15.12.1999 conducted from $16^{40}$ till $17^{15}$, a burst in luminous emittance of plasma generator discharge was observed at $16^{55}$. Therewith the angle θ measured from an arbitrary direction H shown in Fig.6 was equal to 165°, and the value of emittance was 25.7% above its average over the duration of the experiment with a root-mean-square error of ±3.7%. In the experiment of 20.01.2000 performed from $15^{35}$ till $16^{10}$, the burst of plasma luminous emittance was detected at θ = 135° (measured from that same direction H in Fig.6) with 24% excess over the average value of emittance at a root-mean-square error of ±3.3%.

For qualitative understanding of the result obtained and the processing procedure, in Fig.6 some projections of the plasma generator axis on the ecliptic plane corresponding to maximum luminous emittances of plasma discharge are shown by arrows with indication of concrete positions of the Earth in the process of its orbiting around the Sun, the data of the experiments, and the points in time at which the said maximum (with values greater than the experimental error) were observed, The dotted line denotes secondary extremum directions for the emittance. In all experiments the plasma generator rotated counter-clockwise if the plane of rotation seen from above. The direction of the axis of the plasma generator at θ = 0 is indicated by letter H for each experiment.

At the center of Fig.6 (at the site of the Sun) a circle diagram summarizing the results all experiments, is given. They were processed in the following manner. The circle was divided in ten-degree sectors so that the radius-vector passing from the center of the circle along the initial boundary of the first sector was aimed at the point of vernal equinox (21.03) from which the angular coordinate α is counted anticlockwise in the second equatorial system.

In Fig.6 the heights of crosshatched triangles with a 20°-angle at the center of the circle are proportional to the sums (in percentage) of extremum deflections of the oscillograph beam from its average coordinate which out above the standard error of measurements and fall within one or the other of the triangles, for all bursts in luminous emittance observed in all 32 experiments. As is seen from Fig.6, the maximum emittances were observed (more often and with maximum amplitudes) in the $25^{th}$, $26^{th}$ and $33^{th}$, $34^{th}$ sectors.

Notice that when the current vectorial potential of the plasma generator is directed exactly toward $A_g$, the change in $A_\Sigma$ should be maximum and, hence, the magnitude of the new force should be zero since $\frac{\partial \Delta A}{\partial x} = 0$. It follows herefrom for the axially symmetric problem considered that the direction of the vector $A_g$ that the direction of the vector $A_g$ must be related with the sectors 29,30 (Fig.6). This direction has the coordinates: α=290°±10°, and an efficient angle between the axial current of the plasma generator and the vector $A_g$ being equal to 140°±10°.

The result obtained fully coincides (with an error above indicated) with that in which a stationary plasma generator positioned on a special rotatable base and a copper measuring tube with water passed through as a sensing element located in the plasma jet, were used [18,21]. The results of the present experiment are much more precise.

It should be noted that a considerable number of bursts fall into the sectors 11 and 12, corresponding to the direction precisely opposite to that of $A_g$. This may be attributed to the action of back currents in the discharge of the plasma generator directed at an angle φ ≈ 30° to its axis.

From 10.05.2000 till 31.05.2000 and from 11.10.2000 till 3.11.2000, a run of experiments was carried out with scanning the celestial sphere in the vicinity of sectors from $25^{th}$ till $34^{th}$ for determining most efficient angles of special position of the plasma generator axis relative to the vector $A_g$, i.e. the angles of maximum action of the new force.

The investigation has given the following results. It was brought to light that the new force rejects substance out of the region of weakened summary potential $A_\Sigma$ along the generatrixes of a cone with an opening of 100° around the cosmological vectorial potential $A_g$ having the coordinates α ≈ 293° ± 10°, δ ≈ 36° ± 10° in the second equatorial coordinate system. Today it is more accurate result for the $A_g$ direction.

The direction found qualitatively coincided with the results of earlier experiments and as if generalized them all because three effective directions detected (see Figs. 4) are just generatrixes for the cone or its axis.

Analysis of experimental errors is presented in [18,21,22].

**Manifestations of new force in Astrophysics:**

If the new force really exists, it should be manifest itself in natural phenomena since the realizes all its possibilities.

**Pulsars as sensors for the new force.**

The sources of periodic pulsed radiation in the starry sky (pulsars) were discovered in 1968 at a radioastronomic observatory of Cambridge (Great Britain). The first radiopulsar radiated high-stable pulses with period of 1.337s. To day more than 1000 pulsars are known.

Pulsars are astonishing remnants after gravitational collapses of Super Novas and their subsequent explosions. In the framework of existent physical conceptions one has not succeeded in unique (in the context of anyone only model) explanation of the nature of pulsars, enormous values of their velocities of motion (more than 500 km·s$^{-1}$) as well as an asymmetry in substance bursts in the process of explosion of a Super Nova [29-32], etc.

In the connection with the fact that the pulsars are neutron stars with the characteristic dimension around of 10km and very strong magnetic field B (close to $10^{12}$Gs), these objects become unique probes for investigating the structure of surrounding physical space, physical vacuum (the lowest energy state of physical fields) and the new force. If we multiply the indicated value of field B by the radius of pulsar, we obtain the magnitude of vector potential many orders of magnitude higher than the modulus of $A_g$ (Such an estimation is not correct because is made with the use of the existent electromagnetic theory in which the potentials are physically unobservable quantities. But this estimate characterizes the scale of events). As the magnitude of a vector potential cannot be larger than the modulus of $A_g$, all basic properties of the new anisotropic force are bound to clearly manifest themselves in the vicinity of a pulsar. Let's show that. At the collapse of a Super Nova, when the process of generation of enormous magnetic fields is going, the substance must be rejected by the new force out of the region of collapse and decreased summary potential $A_\Sigma$. Due to the action of the reactive force the pulsar must acquire a potent pulse to the side opposite to the direction of the new force. Now let's see what a statistical analysis of real data for pulsars shows.

In my common works with A.A. Shpitalnaya and I.F. Malov is shown [13,14]:

- In the context of the accepted traditional physical models of pulsars, it is quite impossible to explain their enormous velocities up to 1000 km·s$^{-1}$. But these magnitude fall in the range from 10km·s$^{-1}$ to tens of thousand km·s$^{-1}$ obtained in accordance with the byuon theory and explained there by the action of the new anisotropic force.

- According to all existent views of isotropicity of space, the pulsars must move randomly in the sky without any chosen direction. In our work [13] is shown that the distribution of tangential velocities of pulsars (velocities of their motion over the starry sky) convincingly brings to light an asymmetry in their arrangement (Fig.7). At that the bulk of pulsars have tangential velocities directed oppositely to the vector $\mathbf{A_g}$, which corresponds to the action of the new force (more exactly, the reactive force). The latter is strengthened by the fact that the tangential velocities of pulsars are zeroth along the line of vector $\mathbf{A_g}$, and a considerable portion of these velocities are directed at angles of 50° - 60° relative to the reverse direction to $\mathbf{A_g}$, that is, they are lying lie on the reverse cone of action of the new force. The last our work with I.F. Malov [14], where 147 pulsas from the catalog [32] were considered, only considerably strengthened this result.

- By our work, the dependence of pulsar's velocities on magnitudes of their magnetic fields is proved. The normal pulsars with the field about of $10^{12}$Gs have an average velocity close to 430 km·s$^{-1}$ whereas for the millisecond pulsars with the field of $10^9$Gs the magnitude of average velocity is about of 138 km·s$^{-1}$. This conclusion is also confirmed by basic properties of the new force.

- The results obtained allow to hope that the new interaction will be helpful for solution of problems associated with physical processes in pulsars and around them, and will be useful for further understanding of the nature itself of pulsars.

**The dark energy and the new force.**

Let us discuss the nature of the dark energy in the framework of the new force action on the base of potentials of physical fields. It is known that the gravitational potential φ is negative, and therefore for any summation of potentials it decreases the modulus of $\mathbf{A_\Sigma}$. Masses of elementary particles are proportional to this modulus [17,18]. Hence the new force will push out any material body from the region of the decreased modulus of $\mathbf{A_\Sigma}$, because a defect of energy $\Delta E = \Delta mc^2$ will appear and the corresponding force will act to the region with undisturbed value of $\mathbf{A_\Sigma}$.

In [15] we have considered an interaction of two galaxies with $10^{10}$ stars, assumed that the mass of each star is of order to the solar mass ( $10^{33}$ g ) and a relative velocity of each galaxy V = 100 km/sec and 1000 km/sec. As the result we obtained the distance $R_{GG}$ where the new force F (1) will be higher than the gravitational force $F_g$: $R_{GG} \geq 10^{26}$ cm for V = 100 km/sec; $R_{GG} \geq 10^{28}$ cm for V = 1000 km/sec.

The physics of the result obtained is simple and connected with the fact that the gravitational interaction since with the distance (r) as $1/r^2$, and the new force as associated with the change in potential φ, diminishes as $1/r$. Hence, at large distances the new force will dominate the gravity.

Thus we have estimated the magnitude of the distance between galaxies above which they scatter under the action of the new force. The estimate obtained is perspective to explicate the nature of the dark energy.

Other manifestations of the new force in nature (new principle of motion and Sun's flight to Hercules constellation (apex of the Sun); motion of Sun's and Earth's magnetic poles; the physical nature of earthquakes and starquakes) may be acquainted with in the books [17,18].

**Experimental method with the aid of high′- precision gravimeter.**

An ideology of experimental investigation of new force with the aid of gravimeters was based on formula (1) in the following manner [16-18]. Assume that due to changes in current systems generating the magnetic field of the Earth, the Sun, or any other sources in Galaxy, a change in the summary potential $\Delta A_\Sigma$ has taken place which change in its one or another form, has been transmitted to the location point of the gravimeters. As the scales of the events considered are huge, the values of $\frac{\partial \Delta A_\Sigma}{\partial x}$ for them are small due to very great magnitudes of the quantity $x$ appearing in the denominator of the formula. But the values of $\Delta A_\Sigma$ can therewith be similarly large and even consist a considerable fraction of the modulus of $\mathbf{A}_g$. Therefore, to detect assumed new signals with small magnitudes of $\frac{\partial \Delta A_\Sigma}{\partial x}$ in the vicinity of the gravimeter, the magnet with $\frac{\partial \Delta A_\Sigma}{\partial x}$ = 100 Gs was arranged close to the test body of the device. In this case the magnet played a part of an amplifier of the signals of new nature.

**Experimental method using plasma device.**

The laboratory is located in the basement floor of the building and is shielded from the outer electromagnetic noise. The experimental installation was comprised of a pulsed plasma generator and a system of measuring the plasma radiation. The plasma generator was fed from an energy-storage capacitor 100mF in total capacity with operating voltage up to 5 kV. The total energy accumulated in the capacitor was equal to ~ 1.25 kJ. The battery was charged from a standard high-voltage power source and commutated to the load (pulse plasma generator) with the aid of a trigatron type air spark gape activated by a short (~1ns) high-voltage (~30kV) pulse coming to the air-gap from a trigger circuit

The case of the generator (its outer electrode being anode) was made of thin-walled copper tube 11 mm in external diameter and 100 mm in length. The axial electrode (cathode) 4 mm in diameter made from copper bar was placed into acrylic plastic tube with inner diameter of 4 mm and outside diameter equal to that of the outer electrode. The plasma generator was locked on a textolite plate positioned on a special adjustment table rotatable around its vertical axis.

The table was provided with a limb allowing to control the angle of rotation of the whole system relative to some starting position. The plate itself could rotate around the horizontal axis through an arbitrary angle β. The system as an assembly made it possible to rotate the plasma generator during the experiment around the vertical axis through any angle, and around the horizontal axis through any angle in the range -90°< β<60°. It was assumed that at β<0 the discharge of the plasma generator turns to the surface of the Earth. The horizontal position of the plasma generator corresponded to angle β=0. The trajectory of motion of the face of plasma generator during its rotation in the horizontal plane represented a circle. All experiments were carried out in air at atmospheric pressure.

The amplitude of the current in the first maximum reached 21 kA. The maximum voltage between the electrodes equaled 3.5 kV at 5 kV charging voltage on the energy-storage capacitor.

The axis of the thermodetector was directed to the prior known range of maximum discharge glow lying on the axis of the plasma generator ~ 2cm from its face. .

According to Formula (1), the new force is proportional $I^2$ (since $\Delta A_\Sigma \sim I$, where I is the discharge current). Therefore an increase of current and velocity of particles [18] must lead to growth of additional energy of particles in the discharge and hence to an additional heating of plasma. Because the radiative energy of plasma is proportional to $T^4$, an insignificant change in temperature T could be sensed by the thermodetector.

**Theoretical method**

The theory of byuons [17-20] is theory of "life' of special mathematical discrete objects - byuons from which the surrounding space and the world of elementary particles form. An essential distinction of the byuon theory from modern models in the classical and quantum field theories [23] is that the potentials of physical fields (gravitational, electromagnetic, asf.) become, in the said theory exactly fixable, measurable values. In the existing field theory, potentials are defined only with a precision of an arbitrary constant or the rate of change of the potentials in space or time. But in the theory of byuons, field potentials become single-valued since there are formed, on the set of byuons, field charge numbers which generate the fields themselves, as, for example, the electric charge of an electron generates an electric field.

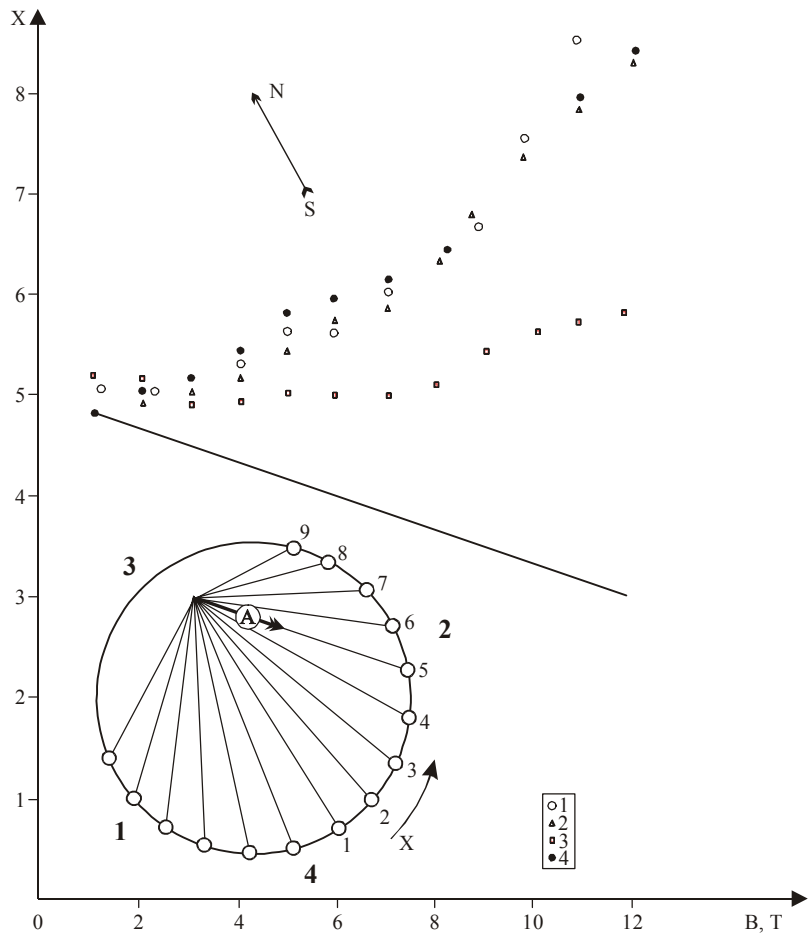

Fig. 1. The deflection *x* of the weight depending on magnetic flux density *B* in the regions *1 ÷ 4* of the solenoid aperture; March 02, 1990,
*t=13h 45m (1), 13h 00m (2), 14h 15m (3), 15h 00m (4)*;
the straight line is a linear approximation (theory);
*A* is the weight from *β*-tin.

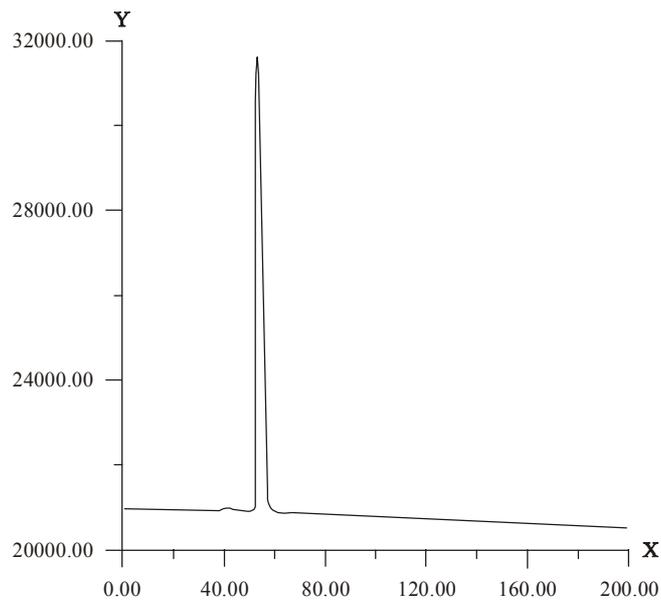

Fig. 2. The event on March 18, 1996.
*y* is displacement of platinum weight. One division is *0.1 μm*, or *0.2 μgal*, *x* is time in minutes.

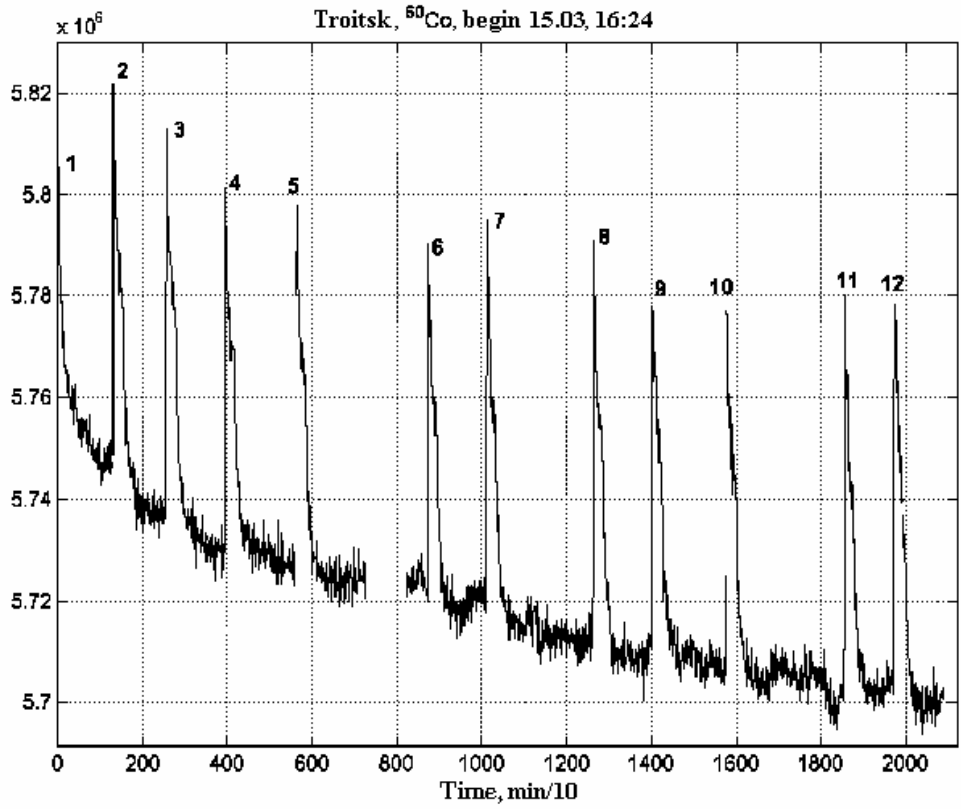
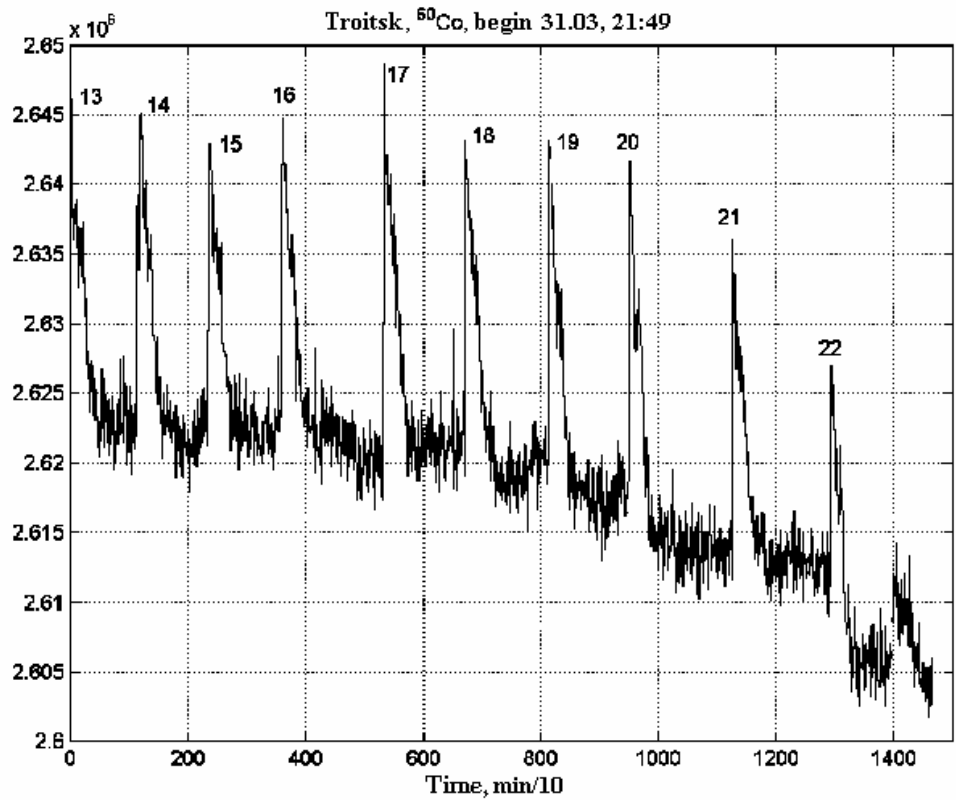

Fig. 3. The variation of the flow of γ-quanta accompanying the β-decay of $^{60}Co$, with the time (INR, Troitsk).

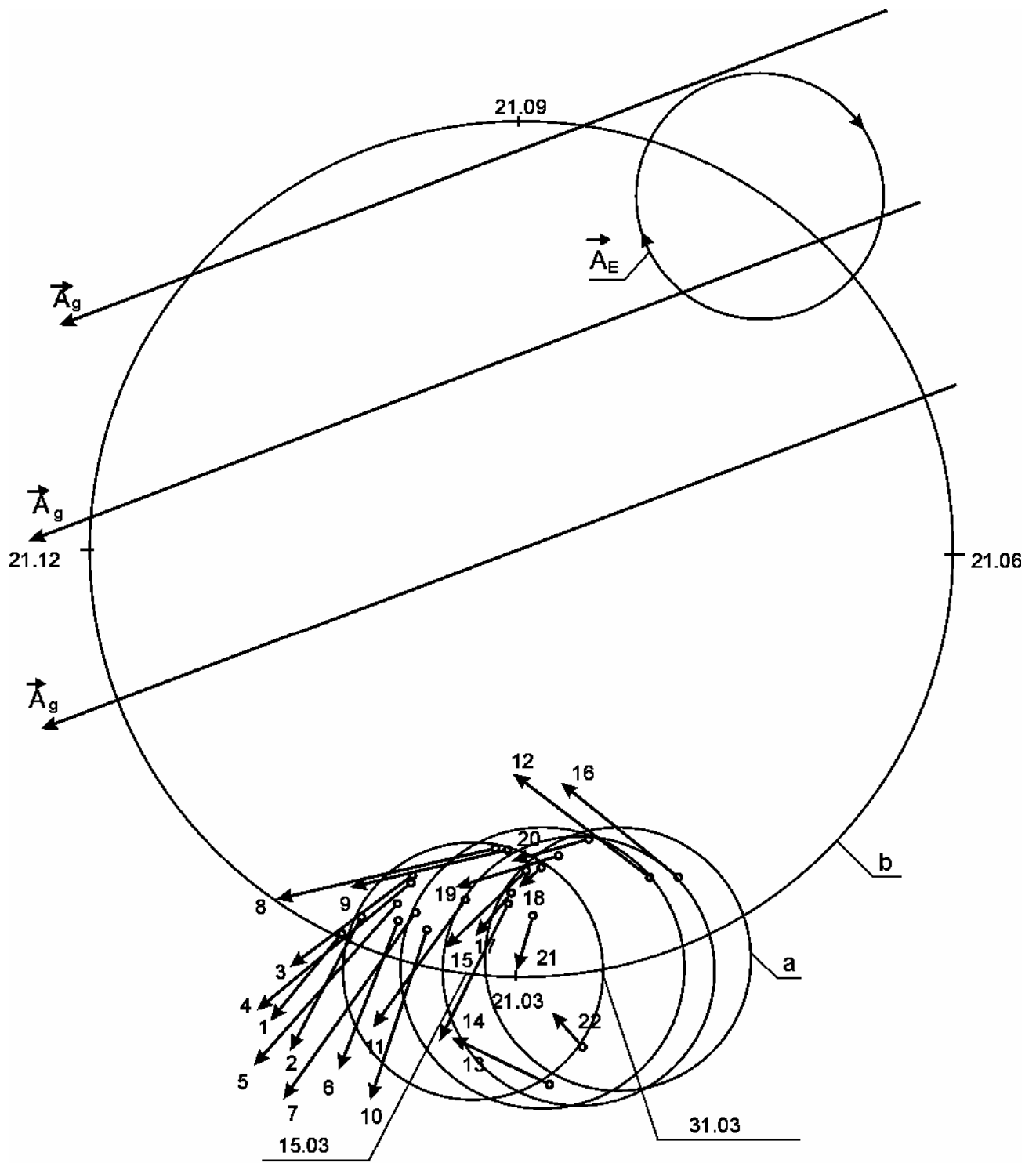

Fig. 4. The spatial positions of sites where the clearly expressed extrema in the magnitude of the flow of γ-quanta in the experiment with the β-decay of $^{60}$Co, were observed (see Fig.3).
←• - the site of the maximum flow of γ-quanta with the indication of the direction of action of the new force drawn along the tangent line to the parallel of latitude;
a – the trajectory of motion of the radioactive source rotating together with the Earth
b – the trajectory of motion of the Earth and the radioactive source around the Sun;
21.03 etc. –the point of the vernal equinox and other characteristic points of the trajectory "b";
$A_E$ –the direction of the vectorial potential of the magnetic field of the Earth's dipole;
$A_g$ – the direction of the cosmological vectorial potential.

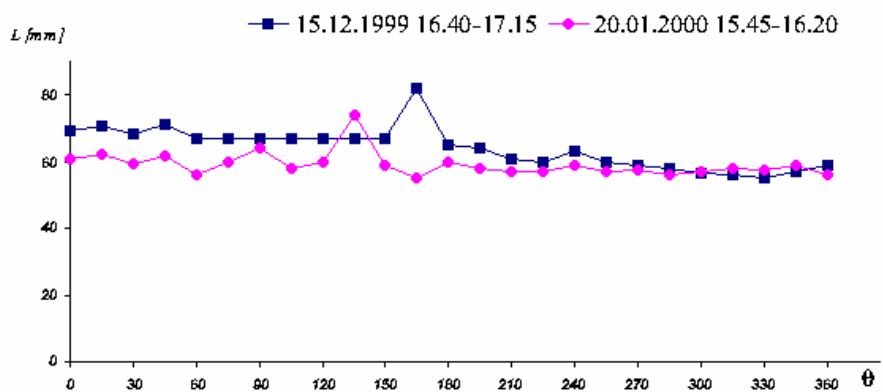

Fig. 5. The magnitude L (mm) of beam deflection of the mirror-galvanometer oscillograph in dependence of angle of rotation θ for the experiments of 15.12.1999 (16.40..17.15) and 20.01.2000 (15.45..16.20).

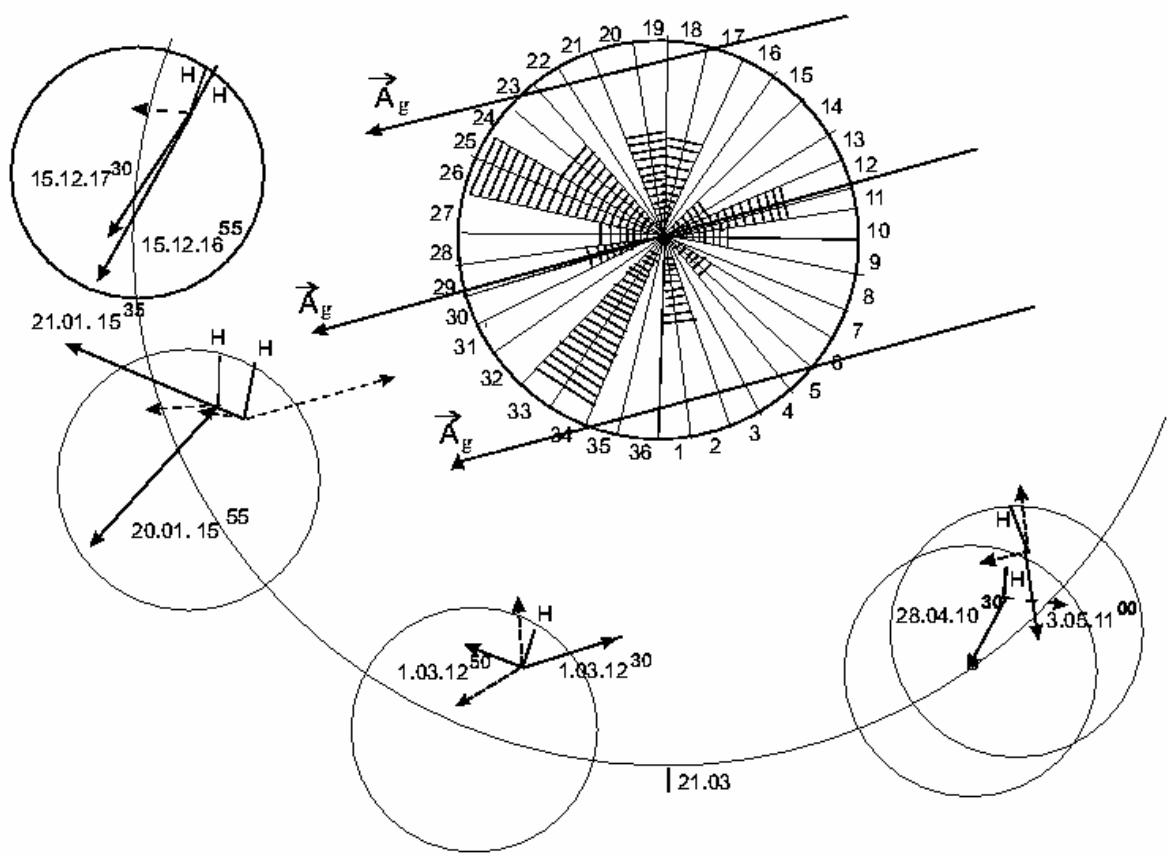

Fig. 6. Directions of the axis of the plasma generator along which maximum deflections of the oscillograph beam (indicated by arrows) when rotating the plasma generator in the horizontal plane. By H a start of rotation.(θ = 0) is denoted.
Indicated are data and Moscow times of observation of maximums in beam deflection.
At the center of the Figure, heights of cross-hatched triangles correspond to the sums of beam deflection magnitudes (in percent) from their average value exceeding the error (for a given sector).
$A_g$ is the cosmological vectorial potential.

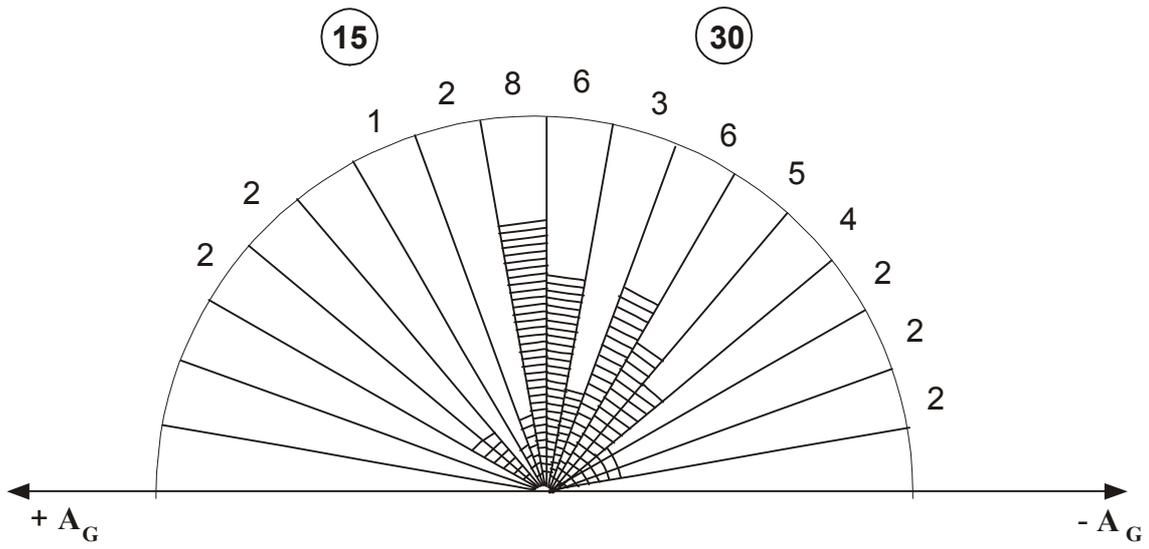

Fig. 7. The results of computation of spatial distribution of the vectors of pulsar tangential velocities $V_t$ (motion velocities in the plane of the sky).